# Maximum efficiencies and performance limiting factors of inorganic and hybrid perovskite solar cells

Yoshitsune Kato, Shohei Fujimoto, Masayuki Kozawa and Hiroyuki Fujiwara*

Department of Electrical, Electronic and Computer Engineering, Gifu University, 1-1 Yanagido, Gifu 501-1193, Japan.

Abstract

The Shockley and Queisser limit, a well-known efficiency limit for a solar cell, is based on unrealistic physical assumptions and its maximum limit is seriously overestimated. To understand the power loss mechanisms of record-efficiency cells, a more rigorous approach is necessary. Here, we have established a formalism that can accurately predict absolute performance limits of solar cells in conventional thin film form. In particular, a formulation for a strict evaluation of the saturation current in a nonblackbody solar cell has been developed by taking incident angle, light polarization and texture effects into account. Based on the established method, we have estimated the maximum efficiencies of 13 well-studied solar cell materials [GaAs, InP, CdTe, a-Si:H, CuInSe$_2$, CuGaSe$_2$, CuInGaSe$_2$, Cu$_2$ZnSnSe$_4$, Cu$_2$ZnSnS$_4$, Cu$_2$ZnSn(S,Se)$_4$, Cu$_2$ZnGeSe$_4$, CH$_3$NH$_3$PbI$_3$, HC(NH$_2$)$_2$PbI$_3$] in a 1-μm-thick physical limit. Our calculation shows that over 30% efficiencies can be achieved for absorber layers with sharp absorption edges (GaAs, InP, CdTe, CuInGaSe$_2$, Cu$_2$ZnGeSe$_4$). Nevertheless, many record-efficiency polycrystalline solar cells, including hybrid perovskites, are limited by open-circuit voltage and fill-factor losses. We show that the maximum conversion efficiencies described here present alternative limits that can predict the power generation of real-world solar cells.

*fujiwara@gifu-u.ac.jp

# I. INTRODUCTION

More than a half century ago, Shockley and Queisser developed a new theory that predicts the ultimate limit of solar cell conversion efficiency [1]. This physical model is based on a simple and straightforward assumption; one photon generates one electron and hole pair in a p-n junction solar cell at thermal equilibrium. Remarkably, the Shockley–Queisser efficiency limit (SQ limit) is estimated by considering only one physical quantity, namely, the band gap ($E_g$) of light absorber at room temperature.

In the last 50 years, the SQ limit has been adopted quite extensively as an absolute criterion that sets the maximum possible limit of the solar cell efficiency and, quite often, the performances of world-record solar cells are compared with those defined by the SQ limits [2–4]. However, despite the substantial research efforts that continued for the last five decades, the conversion efficiencies of the record-efficiency single cells ($\eta \leq 29.1\%$ in Ref. [5]) are still significantly inferior to the maximum SQ efficiency of 34% obtained at $E_g \sim 1.4$ eV [2–4].

Nevertheless, the lower conversion efficiencies observed in experimental cells are, in part, due to the overestimation of the efficiency limit in the SQ model, which is based on unrealistic physical assumptions that can never be achieved in the real world. Specifically, the SQ theory assumes infinite thickness of the light absorber with absolutely zero light reflection (i.e., a perfect blackbody absorber). The zero light reflection occurs only when the refractive index ($n$) of the p-n diode is one with no light absorption [i.e., extinction coefficient ($k$) is zero], whereas conventional absorbers show $n = 3 \sim 4$ with $k > 0$. Thus, the SQ model is highly hypothetical in the optical point of view. Because of its simplicity, this model does not account for the unfavorable parasitic light absorption induced by transparent conductive oxide (TCO) and rear metal electrodes. In the model, the shadow loss (~5%) caused by the front metal-grid electrode is also neglected. Under the simple assumption of the SQ model, the short-circuit current density ($J_{sc}$) is seriously overestimated compared with experimental solar cells.

A step-function variation of light absorption at $E = E_g$, assumed in the SQ model, further contributes to overestimating the open-circuit voltage ($V_{oc}$) [6–8]. In particular, almost all the semiconductor materials show a finite absorption tail owing to the presence of the tail states [9]. Some studies have already pointed out that the tail absorption deteriorates $V_{oc}$ rather significantly owing to the increase in the saturation current density $J_0$ [6,8]. In fact, quite a clear correlation between the $V_{oc}$ loss and the tail absorption has already been reported in experimental solar cells [10].

Accordingly, the SQ limit is expected to show substantially higher efficiencies compared with experimental solar cells that have vital limitations, including finite absorber thickness and a nonideal absorption edge. To critically understand the performance limitation of record-efficiency solar cells, a more rigorous evaluation method that can replace the simple SQ model is necessary. So far, to determine the limiting efficiencies of thin-film-based solar cells, only simple calculations have been performed [6,11–13].

In this paper, we report the development of a rigorous approach that incorporates all the physical and optical aspects of real-world solar cells by advancing the SQ model. In particular, we have evaluated theoretical efficiency limits for a perfectly realizable thin-film solar cell structure (thin-film quantum efficiency limit; QE limit) by adopting true absorption characteristic of the light absorbers. The optical confinement effects due to texture, antireflection coating and backside reflection are fully incorporated into our model. To estimate the thermal balance limit in a nonblackbody cell, strict polarization- and angle-dependent calculation was implemented. Our QE-limit evaluations show that over 30% efficiencies can be realized in 1-µm-thick solar cells with light absorbers having sharp absorption edges. Our potential efficiency calculations in the thin-film configuration further allow us to accurately evaluate the performance limiting factors of record-efficiency photovoltaic devices.

## II. PHYSICAL MODEL

### A. Concept of thin-film limit

Figure 1(a) shows a physical model assumed in the SQ limit calculation. In this model, a p-n homojunction cell having infinite thickness is assumed while forcing the reflectance $R$ of this hypothetical structure to zero (i.e., $R = 0$). In our thin-film concept [Fig. 1(b)], we considered a quite general layer-stacked structure of TCO/absorber (p-n layers)/metal (Ag) with a total absorber-layer thickness of 1 µm. In the structure, a dual antireflection coating of $MgF_2/Al_2O_3$ is further considered to suppress $R$ and is optimized for different solar cells (see Supplemental Material Table I [14] for the exact layer thicknesses). In addition, for a front metal grid electrode, a shadow loss of 5% is further assumed.

In general, the presence of the TCO is problematic, because the free carrier absorption in the TCO deteriorates $J_{sc}$ rather significantly [15]. In our model, to suppress the parasitic absorption in the TCO, a high-mobility TCO layer with a mobility of 100 cm$^2$/(Vs) at a carrier concentration of $2 \times 10^{20}$ cm$^{-3}$ is assumed. Such high mobility has indeed been confirmed in H-doped $In_2O_3$ [15]. The optical constants of the component layers in the structure of Fig. 1(b) are taken from Ref. [9], whereas those of the TCO were calculated by assuming the above mobility and carrier concentration using the Drude model [16,17]. The optical spectra used in our calculations and their modeling parameters are summarized in Supplemental Material Figure 1 and Supplemental Material Tables II–IV [14], respectively. For the modeling of the Ag and $In_2O_3$:H dielectric functions, the Tauc-Lorentz model [18,19] has been applied.

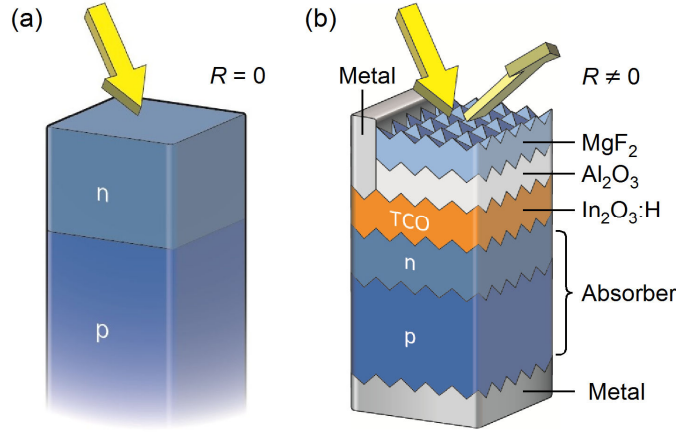

FIG. 1. (a) Optical model adopted for the calculation of the SQ limit and (b) optical model assumed for the thin-film QE limit calculation. In (b), a total absorber thickness is assumed to be 1 μm.

**B. Calculation Method**

In our model, a thin-film solar cell with the structure shown in Fig. 1(b) is placed in a spherical cavity surrounded by a blackbody radiator (Fig. 2) and $J_0$ is estimated by integrating the blackbody radiation (300 K) from all directions inside the cavity with an incident angle $\theta$ and a rotation angle $\phi$. In the case of the SQ calculation, a planer p-n junction solar cell of Fig. 1(a) is placed inside the cavity. In the SQ theory, the solar cell is in a thermal equilibrium, which guarantees that the rate of photon emission is exactly the same as the rate of photon absorption.

The blackbody radiation for a wavelength $\lambda$ at a temperature $T$ is expressed by a well-known equation:

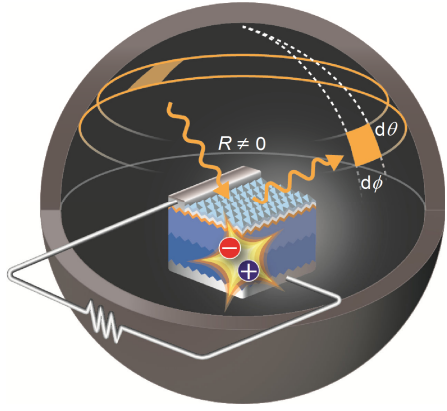

FIG. 2. Physical model for a thin-film solar cell placed in the cavity surrounded by a blackbody radiator.

$$\gamma_{BB}(\lambda,T) = \frac{2hc^2}{\lambda^5}\left[\exp\left(\frac{hc}{\lambda kT}\right)-1\right]^{-1}, \qquad (1)$$

where $h$, $c$ and $k$ are Plank's constant, the speed of light and Boltzmann's constant. The unit of $\gamma_{BB}$ is $Wm^{-2}m^{-1}sr^{-1}$ and the insertion of $T = 300$ K into Eq. (1) gives $\gamma_{BB}$ at room temperature. The density of blackbody photons ($\varphi_{BB}$) with solid angle $d\theta$ and rotation angle $d\phi$ (see Fig. 2) is expressed by

$$\varphi_{BB}(\lambda,\theta)d\theta d\phi = \frac{\lambda \gamma_{BB}(\lambda,T)}{hc}\cos\theta \sin\theta d\theta d\phi. \qquad (2)$$

The $\cos\theta$ in Eq. (2) indicates the projected area on the solar cell surface. In the case of the SQ theory that assumes a perfect blackbody, the integration of Eq. (2) is simplified greatly and the term of $\cos\theta\sin\theta d\theta d\phi$ in Eq. (2) becomes π.

Unlike the SQ theory, in our extended method used for the thin-film QE limit calculation, the effect of light reflection is fully incorporated. Because an imperfect blackbody cell with an exact thin-film structure is assumed, the light absorption and reflection are now incident-angle dependent. This further requires the polarization-dependent calculation for the incident light. Thus, a strict $J_0$ integration needs to be performed inside the cavity. Consequently, $J_0$ of the thin-film solar cells is determined by integrating the blackbody radiation from all the internal surface of the cavity using an exact formula:

$$J_0 = \frac{(1-S)q}{2} \int_0^\infty \int_0^{2\pi} \int_0^{\pi/2} [Q_p(\lambda,\theta) + Q_s(\lambda,\theta)] \varphi_{BB}(\lambda,\theta) d\theta d\phi d\lambda, \quad (3)$$

where $S$ represents the shadow loss of the solar cell ($S = 0.05$) and $q$ shows the electron charge. The $Q_p(\lambda,\theta)$ and $Q_s(\lambda,\theta)$ in the above equation represent the external quantum efficiency (EQE) spectra calculated for the p-polarization and s-polarization at the incident angle $\theta$, respectively. As confirmed from Eq. (3), $J_0$ in this study was evaluated by integrating the EQE response calculated for each angle and p-/s-polarization within the spherical cavity. Because of the non-polarized nature of the blackbody radiation, the EQE response for natural light is estimated as $(Q_p + Q_s)/2$. In Eq. (3), because the solar cell rear surface is covered uniformly with the metal electrode, the rear surface is assumed to be a perfect reflector and the solid angle integral is taken over the hemisphere (i.e., $\theta = 0 \sim \pi/2$) with a step of $0.1^o$.

If $\theta = 0^o$ is assumed in Eq. (3), the distinction between the p- and s-polarization disappears and Eq. (3) is reduced to

$$J_0 = (1-S)q \int Q(\lambda) \varphi_{BB}(\lambda) d\lambda, \quad (4)$$

where $\varphi_{BB}(\lambda) = \pi \lambda \gamma_{BB}/(hc)$ and $Q(\lambda)$ shows the EQE spectrum of a solar cell at normal incidence ($\theta = 0^o$). The above simple equation has been applied widely to estimate approximate $J_0$ values of various solar cells in earlier studies [7,8,20–22]. In this case, however, the EQE calculation is made only at $\theta = 0^o$ and the angler effect in the configuration of Fig. 2 is neglected completely. In this study, therefore, we have also investigated the validity of such an assumption.

On the other hand, $J_{sc}$ of the solar cell is calculated according to a standard equation, which is similar to Eq. (4):

$$J_{sc} = (1-S)q \int Q(\lambda) \varphi_{sun}(\lambda) d\lambda, \quad (5)$$

where $\varphi_{sun}(\lambda)$ indicates the photon density for the solar irradiance under AM1.5G illumination. Once $J_{sc}$ and $J_0$ are obtained from the above procedures, the $J$-$V$ curve of the corresponding solar cell can be obtained according to a standard formula:

$$J = J_0 \left[ \exp\left(\frac{qV}{kT}\right) - 1 \right] - J_{sc}, \quad (6)$$

from which $V_{oc}$, FF and conversion efficiency are readily obtained. As known well, if $J = 0$ is assumed in Eq. (6), $V_{oc}$ can be obtained directly as

$$V_{oc} = (kT/q)\ln(J_{sc}/J_0 + 1) \quad (7)$$

As a result, if $J_0$ deduced from Eq. (3) is applied, the maximum efficiency under the QE

limit is deduced, while $J_0$ obtained from Eq. (4) with the assumptions of $Q(\lambda) = 1$ ($E \geq E_g$), $Q(\lambda) = 0$ ($E < E_g$) and $S = 0$ leads to the SQ limit. It should be noted that, in the above theoretical calculation, the non-radiative recombination is neglected.

**C. EQE Calculation**

In thin-film based solar cells, the optical confinement is critical. In particular, an antireflection effect generated by surface textures is of significant importance to suppress the front light reflection. To fully incorporate all the possible optical effects generated in textured thin-film solar cells, the EQE was calculated based on the ARC approach [23]. In this method, the $R$ spectrum of a textured structure ($R_{tex}$) is first evaluated by forcing antireflection conditions (ARC) in the calculation of a flat optical model and, by applying this $R_{tex}$, the absorptance spectrum (i.e., EQE spectrum) of a textured light absorber is then estimated while assuming a flat layered structure. This quite simple method has been applied successfully to reproduce the optical response of numerous submicron-textured experimental cells, including $CuInGaSe_2$ (CIGSe) [23], $Cu_2ZnSnSe_4$ (CZTSe) [24], $Cu_2ZnSnS_4$ (CZTS) [24], $CuZnSn(S,Se)_4$ (CZTSSe) [24], CdTe [24], $CH_3NH_3PbI_3$ (MAPbI$_3$) [24,25] and $HC(NH_2)_2PbI_3$ (FAPbI$_3$) [26] solar cells. For the EQE calculations based on the ARC method, a software can be used [27].

Figure 3(a) shows the example of the ARC calculation performed for the CdTe solar cell with the structure shown in Fig. 1(b), assuming $\theta = 0°$. The $R_{flat}$ in Fig. 3(a) indicates $R$ obtained assuming a flat structure and the interference fringe appears in $R_{flat}$. In the ARC approach, the minimum $R$ points in the $R_{flat}$ spectrum are linearly connected to express the antireflection effect of the texture. By employing the resulting reflectance spectrum ($R_{tex}$), the corresponding EQE spectrum is estimated assuming 100% carrier collection in the absorber layer.

Once the EQE spectrum is obtained, $J_0$ and $J_{sc}$ can further be obtained. Figure 3(b) shows the EQE spectrum of the calculated CdTe cell, together with the photon density of the blackbody radiation at 300 K ($\varphi_{BB}$). As described in Eq. (4), the approximate $J_0$ value can be obtained by integrating the product of $Q(\lambda)$ and $\varphi_{BB}(\lambda)$. Because $\varphi_{BB}$ increases at low $E$, $J_0$ is primarily determined by the EQE response in the longer $\lambda$ region. As a result, the longer-$\lambda$ EQE response is directly linked to the maximum $V_{oc}$ obtainable through Eq. (7). In our thin-film QE limit calculation, $J_0$ is not determined from a single EQE spectrum obtained at $\theta = 0°$ but is evaluated strictly by performing the exact integration of the blackbody radiation shown in Fig. 2. In the calculation of $Q_{p,s}(\lambda,\theta)$ using the ARC method, the small spikes that appear in the calculated $R$ spectra

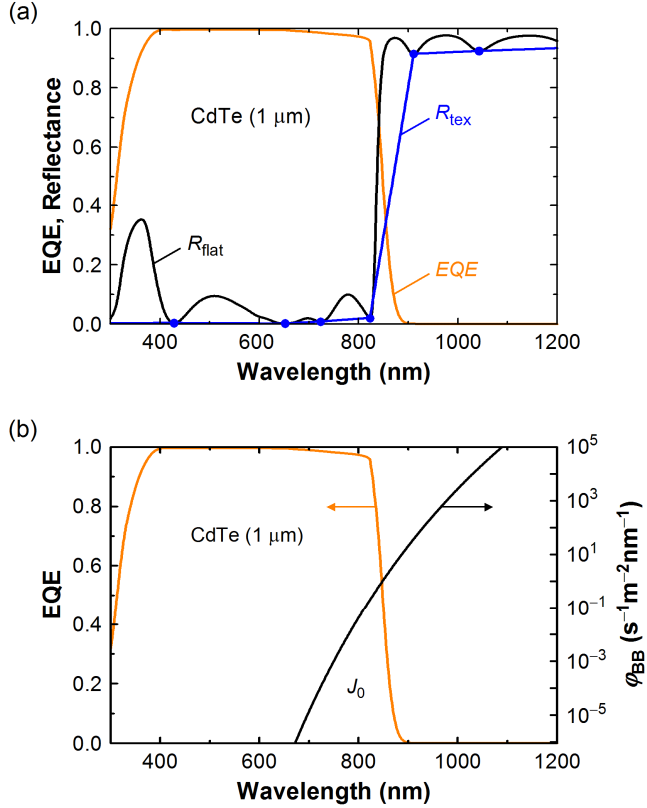

FIG. 3. (a) EQE calculation and (b) $J_0$ calculation performed for the CdTe solar cell assuming $\theta = 0°$. In (a), the black line indicates $R$ obtained using the flat optical model ($R_{flat}$), whereas the blue line indicates $R$ estimated assuming a textured structure ($R_{tex}$). The $R_{tex}$ is obtained by linearly connecting the minimum $R$ points of $R_{flat}$ (blue circles). The EQE spectrum is calculated by adopting $R_{tex}$. In (b), the EQE spectrum of the CdTe solar cell and the photon density of the blackbody radiation at 300 K ($\varphi_{BB}$) are shown. The approximated $J_0$ value of the solar cell is derived as the area surrounded by these two spectra.

were removed to obtain smooth-varying EQE spectra.

In our EQE calculation, an absorber thickness of 1 μm is assumed. For the indirect-transition crystalline Si (c-Si) solar cell, however, the calculation was implemented by assuming an absorber thickness of 150 μm. For this calculation, the continuous phase approximation method [28] has been further applied to reproduce the incoherent optical response in a thick c-Si wafer. It should be noted that standard c-Si solar cells are made using a very large pyramid-type texture (~10 μm) [29] and its structure is completely different from the thin-film structure assumed in Fig. 1(b). In this study, the calculation of the 150-μm-thick c-Si solar cell has been implemented based on the ARC approach to provide a complementary reference for comparison with other thin-film-based solar cells. For the accurate characterization of pyramid-textured c-Si solar cells, more exact approaches [29,30] are necessary.

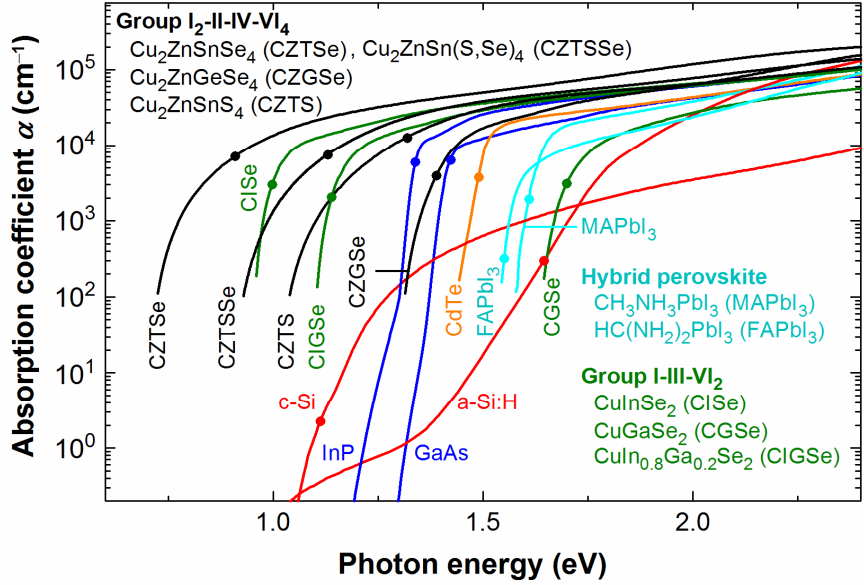

FIG. 4. $\alpha$ spectra of solar cell absorbers. The closed circles indicate the $E_g$ positions of the absorbers. The optical data were taken from Ref. [9], except for CZTSSe.

**D. Absorption spectra of light absorber**

Figure 4 summarizes the absorption-coefficient ($\alpha$) spectra of 14 inorganic and hybrid perovskite absorber materials used for the calculations. All the $\alpha$ spectra were determined by high-precision characterization based on spectroscopic ellipsometry [9] but the $\alpha$ spectra in the low $\alpha$ region ($\alpha < 100$ cm$^{-1}$) have been determined by further combining other complementary techniques including transmission measurements. Moreover, the $\alpha$ spectra of the alloy semiconductors (CIGSe and CZTSSe) were derived based on the reported dielectric-function modeling methods [24,31]. Although controversial optical data have been reported for CIGSe [32] and hybrid perovskites

Table I. Comparison of the assumptions in the QE limit with those in the SQ limit.

| Model parameters | QE limit | SQ limit |
|---|---|---|
| Absorber thickness | 1 μm (variable) | ∞ |
| Reflectance | $R \neq 0$ | $R = 0$ |
| Temperature | 300 K | 300 K |
| Absorber $\alpha$ | Experimental data | Step function at $E_g$ |
| Tail absorption | $E_U > 0$ eV | Neglected ($E_U = 0$ eV) |
| Optical model | Textured | Flat |
| Shadow loss | 5% | Neglected |
| Front TCO absorption | Considered | Neglected |

[25], this controversy is found to originate from the effect of rough surface [33] and quite consistent optical characterization has been made to minimize the roughness effect in the result of Fig. 4. Rather surprisingly, all the direct-transition semiconductors show similar $\alpha$ values of ~$10^4$ cm$^{-1}$ in the band-edge region.

The closed circles in Fig. 4 represent the $E_g$ positions. Many single and polycrystalline materials exhibit sharp absorption edges with Urbach energies of $E_U$ ~ 10 meV [9]. Nevertheless, in CZTSe, CZTS, CZTSSe and hydrogenated amorphous silicon (a-Si:H), quite strong tail-state absorption is confirmed at $E < E_g$ owing to the presence of extensive disorder. The tail state formation in a-Si:H is caused by the random nature of the amorphous network [34]. Quite large tail absorption in CZTSe and CZTS has been attributed to cation disorder (i.e., Cu/Zn/Sn mixing) [35,36], while the tail state generation is negligible in Cu$_2$ZnGeSe$_4$ (CZGSe) due to the suppression of the cation mixing [36].

**E. Comparison between the QE and SQ limits**

Table I summarizes the difference of the fundamental assumptions between the QE and SQ approaches. As mentioned above, a critical feature of the QE limit is a finite absorber thickness that generates a non-zero $R$, while the unrealistic infinite absorber thickness with $R = 0$ is assumed in the SQ model. Another large improvement of the QE model over the SQ model is the inclusion of the absorber tail absorption, which has been incorporated into the calculation by adopting experimental data, while an unrealistic step function with $E_U = 0$ eV is assumed for the light absorption in the SQ model. In the case of the QE limits, textured effects have been considered using the ARC method, whereas the $J_{sc}$ reduction induced by the shadow loss and the TCO parasitic absorption has also been incorporated explicitly based on the optical model shown in Fig. 1(b).

**III. RESULTS**

**A. Simulated EQE spectra**

Figure 5 summarizes the EQE spectra calculated by applying the ARC method ($\theta = 0°$) using the thin-film structure shown in Fig. 1(b) and the $\alpha$ spectra shown in Fig. 4. The closed circles in Fig. 5 represent the $E_g$ positions of the absorber layers. Despite a rather thin absorber thickness assumed in the calculation, near 100% EQE is realized at $E > E_g$ by fully incorporating texture and antireflection-coating effects. The reduction of EQE in the short $\lambda$ region (< 400 nm) is caused by the TCO parasitic absorption, whereas the longer $\lambda$ limit is determined by the light absorber. When the absorber has extensive tail states, the corresponding EQE spectrum shows a noticeable EQE tail. This effect is quite significant in the disordered materials (i.e., CZTSe, CZTS, CZTSSe and a-Si:H).

From the EQE spectra shown in Fig. 5, $J_0$ can be calculated using Eq. (4). We confirmed that the $J_0$ calculations under the strict integration using Eq. (3) and the simple calculations using Eq. (4) result in $J_0$ values in similar ranges and there is a clear relationship between the $J_0$ values calculated by Eqs. (3) and (4) (Fig. 6). Accordingly, although the exact calculation using Eq. (3) is preferable, Eq. (4) can still be adopted to estimate an approximated value. In this study, however, accurate $J_0$ values calculated by Eq. (3) have been used to estimate the maximum efficiencies and performance limiting factors.

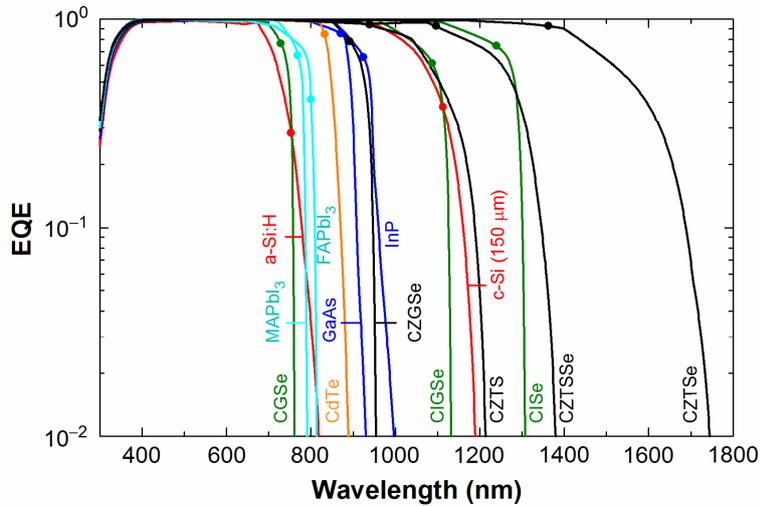

FIG. 5. EQE spectra calculated from the ARC method assuming the optical model of Fig 1(b) using $\theta = 0°$. The closed circles indicate the $E_g$ positions of the absorbers.

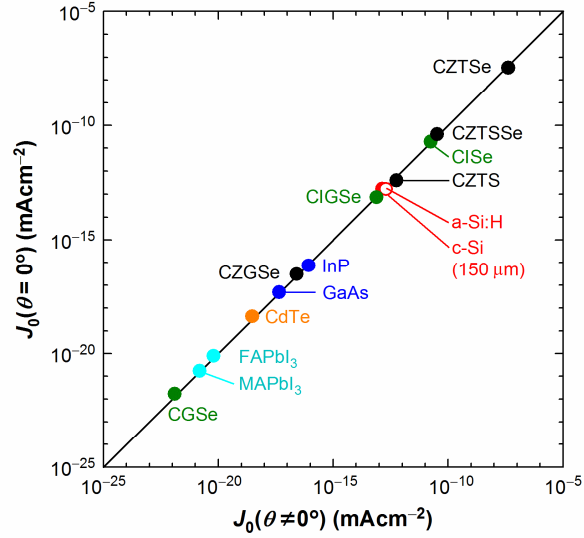

FIG. 6. Relationship between $J_0$ values calculated by assuming $\theta \neq 0°$ and $\theta = 0°$. The $J_0$ values of $\theta \neq 0°$ were evaluated by performing exact spherical integration of Fig. 2 using Eq. (3), whereas the $J_0$ values of $\theta = 0°$ were obtained by using Eq. (4) assuming the normal incidence for the EQE calculation.

**B. Maximum efficiencies**

Figure 7 shows the maximum solar cell parameters obtained assuming the thin-film QE limit (closed circles) and the SQ limit (solid lines). For c-Si, the result for an absorber thickness of 150 μm is shown. The numerical values in Fig. 7, including the maximum $V_{oc}$ ($V_{oc}^{max}$), $J_{sc}$ ($J_{sc}^{max}$) and $FF$ ($FF^{max}$) of the QE limits, are summarized in Table II. In this table, the reported $E_g$ values [9,24–26,32,36–41] are also shown.

In Fig. 7(a), for many solar cells, $J_{sc}$ obtained in the QE limit is slightly lower than that of the SQ limit owing to the presence of (i) shadow loss of the front electrode, (ii) the non-zero $R$ of the device and (iii) parasitic absorption in the front TCO and rear metal electrodes. However, some disordered materials (CZTS and CZTSSe) show notably higher $J_{sc}$, compared with the SQ limit, because the quite strong tail absorption in these materials allows the light absorption in a spectral range even below $E_g$.

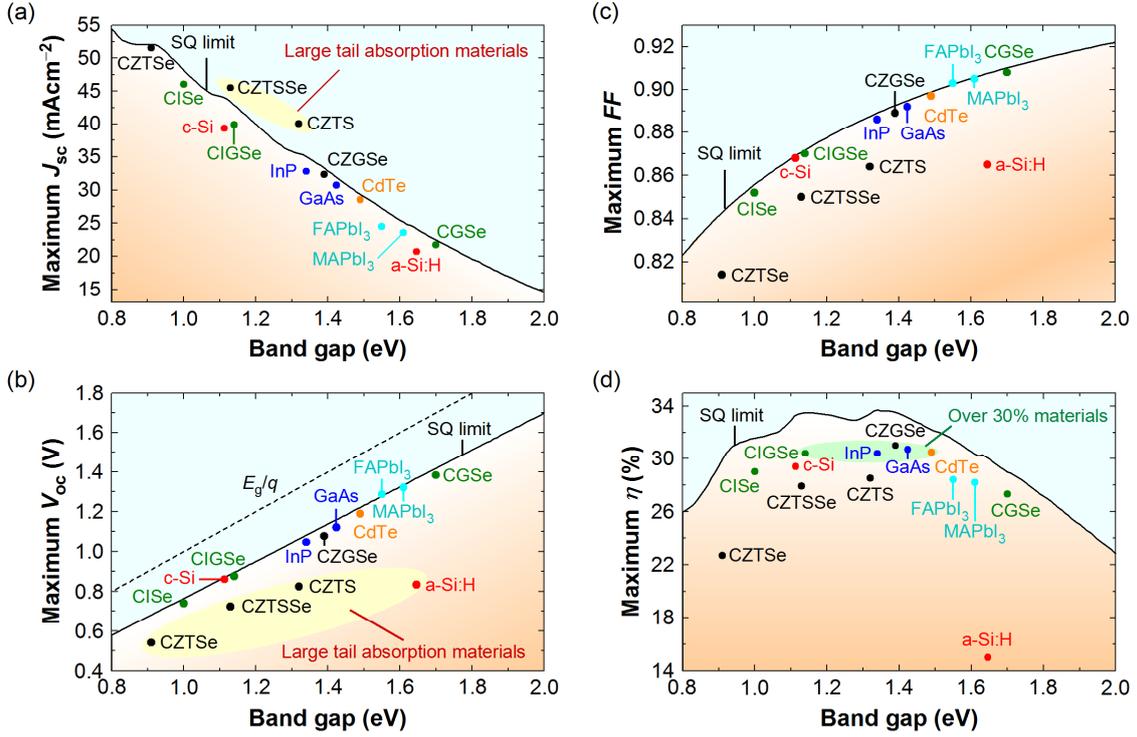

FIG. 7. Maximum (a) $J_{sc}$, (b) $V_{oc}$, (c) $FF$ and (d) conversion efficiency ($\eta$) calculated from the thin-film QE and SQ limits. The closed circles show the calculated QE limits for each absorber, whereas the solid line indicates the SQ limit. The yellow regions in (a) and (b) indicate the absorber materials that exhibit strong tail absorption. In (b), the dotted line indicates $E_g/q$. In (d), the green region indicates the absorber materials that exhibit over 30% efficiencies.

The $V_{oc}^{max}$ obtained in our QE limit calculation is also lower than that of the SQ limit [Fig. 7(b)]. This originates from larger $J_0$ in the QE limit, compared with the SQ limit, and is caused by the incorporation of finite tail absorption in the QE limit calculations. Remarkably, the disordered materials (a-Si:H and CZT(S)Se) exhibit quite low $V_{oc}^{max}$ than that of the SQ limit, owing to the strong EQE tailing in the longer $\lambda$ region (see Fig. 5). As reported earlier [8,42–44], all of these disordered materials suffer from large $V_{oc}$ deficits. Our potential $V_{oc}$ calculation confirms that the large $V_{oc}$ deficit in these disordered phases is caused primarily by the presence of the tail states, which leads to the drastic increase in $J_0$.

Table. II. Maximum conversion efficiencies derived from the thin-film QE and SQ limits. The maximum $J_{sc}$ ($J_{sc}^{max}$), $V_{oc}$ ($V_{oc}^{max}$) and $FF$ ($FF^{max}$) in this table correspond to those of the thin-film QE limit indicated by the closed circles in Fig. 7.

| Solar cell | $E_g$ (eV) | Ref. [a] | $J_{sc}^{max}$ (mAcm$^{-2}$) | $V_{oc}^{max}$ (V) | $FF^{max}$ | QE limit (%) | SQ limit (%) |
|---|---|---|---|---|---|---|---|
| CZTSe | 0.91 | [37] | 51.6 | 0.541 | 0.814 | 22.7 | 29.9 |
| CISe | 1.00 | [32] | 46.0 | 0.739 | 0.852 | 29.0 | 31.6 |
| c-Si (150 μm) | 1.11 | [9] | 39.4 | 0.861 | 0.868 | 29.4 | 33.3 |
| CZTSSe | 1.13 | [24] | 45.5 | 0.723 | 0.850 | 27.9 | 33.5 |
| CIGSe | 1.14 | [32] | 39.9 | 0.876 | 0.870 | 30.4 | 33.5 |
| CZTS | 1.32 | [38] | 40.0 | 0.824 | 0.864 | 28.5 | 33.4 |
| InP | 1.34 | [39] | 32.8 | 1.047 | 0.886 | 30.4 | 33.7 |
| CZGSe | 1.39 | [36] | 32.4 | 1.077 | 0.889 | 31.0 | 33.5 |
| GaAs | 1.42 | [40] | 30.7 | 1.121 | 0.892 | 30.7 | 33.1 |
| CdTe | 1.49 | [41] | 28.6 | 1.189 | 0.897 | 30.5 | 32.2 |
| FAPbI$_3$ | 1.55 | [26] | 24.5 | 1.286 | 0.903 | 28.4 | 31.4 |
| MAPbI$_3$ | 1.61 | [25] | 23.6 | 1.320 | 0.905 | 28.2 | 30.4 |
| a-Si:H | 1.65 | [9] | 20.7 | 0.834 | 0.865 | 15.0 | 30.0 |
| CGSe | 1.70 | [32] | 21.8 | 1.383 | 0.908 | 27.3 | 29.0 |

[a] Reference for $E_g$ of each absorber.

For $V_{oc}^{max}$, if the EQE spectra obtained from experimental solar cells are applied to Eq. (4), $V_{oc}^{max}$ can be obtained empirically. Based on this semi-experimental approach, the $V_{oc}^{max}$ values of 0.86 V (c-Si) [7], 1.146 V (GaAs) [7] and 1.32–1.34 V (MAPbI$_3$) [7,20–22] have been reported. These values are quite consistent with $V_{oc}^{max}$ listed in Table II, confirming the validity of our simulation procedure.

The overall trend of $FF$ is quite similar to that of $V_{oc}$; the QE limits are slightly smaller than the SQ limits and the large $FF$ reduction occurs in the large-tail disordered materials [Fig. 7(c)]. In absolute values, however, the difference in $FF$ between the SQ and QE limits is rather small.

The maximum conversion efficiency in the SQ limit (33.7%) is obtained at $E_g = 1.34$ eV [Fig. 7(d)]. The maximum efficiencies derived from our QE limit calculation are notably smaller than those obtained from the SQ theory due to the limitation of $J_{sc}$, $V_{oc}$ and *FF*. Rather surprisingly, however, even in the 1-μm-thick limit, over 30% efficiencies can still be obtained for CIGSe, InP, GaAs, CdTe and CZGSe, whereas hybrid perovskite solar cells show the maximum efficiencies of ~28% (see Table II) due to their $E_g$ being slightly higher than the optimum of 1.34 eV.

The calculated maximum efficiencies of the disordered materials can be categorized into two group; one group (CZTS and CZTSSe) shows relatively high efficiencies, while very low efficiencies are obtained in the other group (a-Si:H and CZTSe). This can be interpreted by spectral matching; in the high efficiency group, spectral matching between the sun and EQE spectra is good and tail absorption works positively to increase $J_{sc}$. In the low-efficiency group, the $J_{sc}$ gain is not sufficient because of the lack of solar irradiance in the corresponding tail absorption region, making the overall efficiency low. Thus, our calculation is quite valid to estimate the effect of tail absorption on the maximum achievable conversion efficiency.

## C. Effects of absorber thickness and tail absorption

In the above QE limit calculations, the absorber thickness was fixed to 1 μm and the experimental $\alpha$ spectra were applied. To reveal the influence of the absorber thickness and tail absorption explicitly, we have performed the QE limit calculations by varying the absorber thickness and tail state absorption.

Figure 8(a) shows the EQE spectra of GaAs solar cells, obtained when the absorber thickness in the model of Fig. 1(b) is varied in a range of 100 ~ 3000 nm. The EQE calculations were performed using the ARC method and, in the simulations, the thicknesses of the antireflection coating ($MgF_2/Al_2O_3$) were optimized for each absorber thickness. In Fig. 8(a), the longer-$\lambda$ EQE increases significantly as the GaAs absorber thickness increases. This effect originates from lower $\alpha$ values in the band-edge region (see Fig. 4), which limit the total light absorption in the longer-$\lambda$ region.

The thickness dependence of EQE has also been calculated for a hybrid perovskite ($MAPbI_3$) and, from these EQE spectra, the QE limits have been determined. Figure 8(b) summarizes the thickness-dependent QE limits obtained for GaAs and $MAPbI_3$ absorbers. Both solar cells show a rapid increase in efficiency up to the thickness of 500 nm. As confirmed from Fig. 4, $\alpha$ in the $E_g$ region is typically $10^4$ cm$^{-1}$, which corresponds to the light penetration depth ($d_p = 1/\alpha$) of ~1 μm. In the calculation result of Fig. 8(b), however, the optical confinement effect has been incorporated and the QE limit saturates at a thinner thickness of ~500 nm. In other words, for light absorbers with $\alpha \sim 10^4$ cm$^{-1}$ in the $E_g$ region, the thickness of 500 nm is sufficient to achieve high conversion efficiencies.

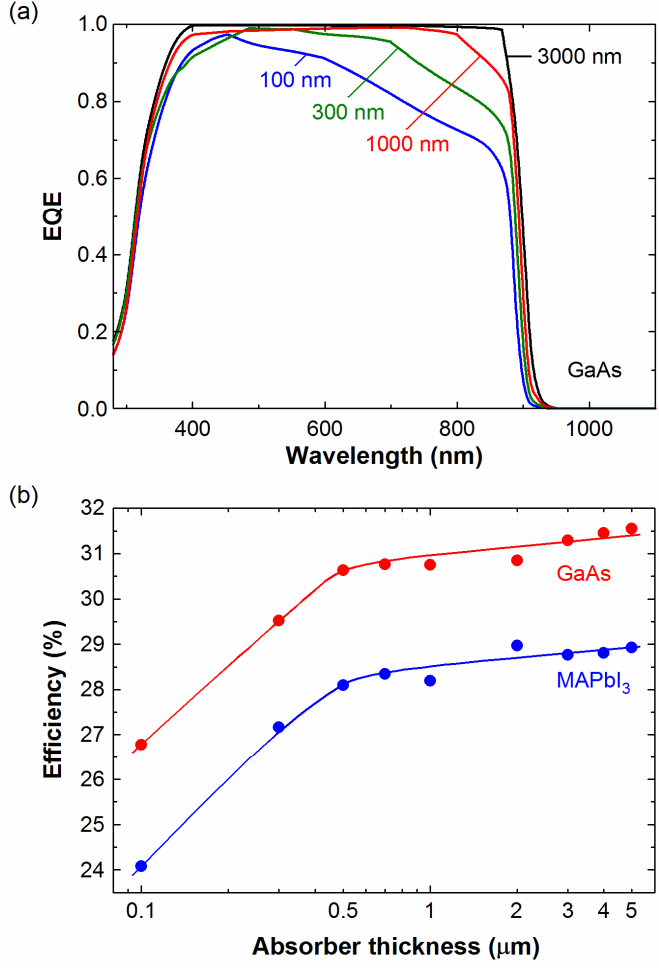

FIG. 8. (a) EQE spectra of GaAs solar cells calculated from the ARC method by varying the absorber thickness in the optical model of Fig. 1(b) and (b) thickness dependence of the QE limit obtained for GaAs and MAPbI$_3$ solar cells.

To determine the effect of the tail absorption on the QE limit, we have considered the hypothetical tail absorption for GaAs, which is modeled based on a simple expression:

$$\alpha(E) = \alpha_{E_g} \exp[(E - E_g)/E_U], \qquad (8)$$

where $\alpha_{E_g}$ denotes the $\alpha$ value at $E_g$. The above model was used only for the energy region of $E < E_g$ and the actual experimental data were applied for $E \geq E_g$. Figure 9(a) shows the result of the $\alpha(E)$ modeling with $E_U$ in a range of 0 ~ 100 meV (solid lines). In the QE limit calculations, the $k$ spectrum calculated from the result of Fig. 9(a) was applied, while the unmodified experimental spectrum was used for $n$. For the QE limit calculations, the identical optical model with a GaAs absorber thickness of 1 μm was applied. Figure 9(b) shows the EQE spectra calculated by the ARC method using different $E_U$ values. It can be seen that the longer-$\lambda$ response increases drastically as $E_U$ increases.

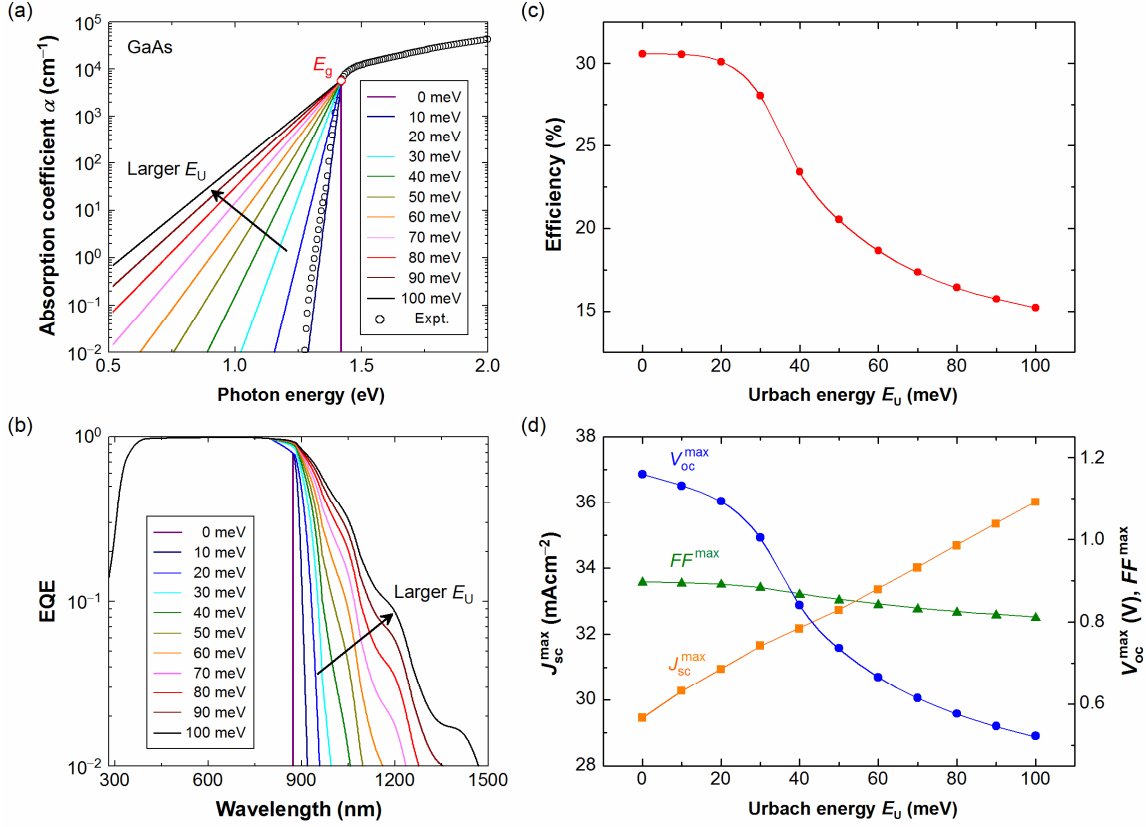

FIG. 9. (a) Modeling of hypothetical GaAs tail absorption by the variation of the Urbach energy ($E_U$) in a range of 0 ~ 100 meV, (b) change of the GaAs EQE spectrum with $E_U$ calculated by the ARC method, (c) variation of the QE limit with $E_U$ and (d) variation of the maximum solar cell parameters with $E_U$. In (a), the open circles show the experimental $\alpha$ spectrum taken from Ref. [9]. The numerical values of (c) and (d) are estimated from the corresponding EQE spectra in (b).

From the EQE spectra of Fig. 9(b), the corresponding QE limits have been evaluated.

Figure 9(c) shows the variation of the QE limit with $E_U$. The $J_{sc}^{max}$, $V_{oc}^{max}$ and $FF^{max}$ obtained from the same calculations are also summarized in Fig. 9(d). In Fig. 9(c), the QE limit is quite high at $E_U \leq 20$ meV but shows a drastic drop at $E_U > 20$ meV, confirming the significant detrimental effect of the tail absorption on the conversion efficiency. Although the larger $E_U$ improves $J_{sc}^{max}$ slightly, the drastic decrease of $V_{oc}^{max}$ with $E_U$ leads to the quite rapid efficiency reduction. When $E_U$ increases, therefore, the overall efficiency is governed by the low $V_{oc}$, caused by the rapid $J_0$ increase due to the tail state absorption, and the optical gain by the tail absorption is not high enough to compensate the $V_{oc}^{max}$ reduction. The result of Fig. 9(c) shows clearly that there is a boundary for efficiency at $E_U = 20$ meV and the light absorber with small $E_U$ ($\leq 20$ meV) is of significant importance to realize high performance.

**D. Limiting factors of record-efficiency cells**

Based on our thin-film limit calculation, the limiting factors of record-efficiency cells are evaluated. For this calculation, the solar cell structure of Fig. 1(b) is adopted but the absorber thickness is adjusted to those of the actual devices [45–55]. Figure 10 summarizes the calculated maximum conversion efficiencies in the thin-film QE limit ($\eta_{QE}^{max}$), together with the conversion efficiencies of the experimental record-efficiency cells ($\eta_{ex}$) reported in Refs. [5,46,50,52,53,55,56]. In Table III, the absorber thicknesses adopted in the actual calculations and the numerical values of $\eta_{QE}^{max}$ and $\eta_{ex}$ are summarized. In Fig. 10, the difference between the theoretical and experimental efficiencies is further categorized as the efficiency reduction $\Delta\eta$ caused by $V_{oc}$ deficit ($\Delta\eta^{Voc}$), $J_{sc}$ deficit ($\Delta\eta^{Jsc}$) and $FF$ deficit ($\Delta\eta^{FF}$) according to the calculation results of $V_{oc}^{max}$, $J_{sc}^{max}$ and $FF^{max}$. In Table III, we also show $V_{oc}$, $J_{sc}$ and $FF$ deficits expressed by $\Delta V_{oc} = V_{oc}^{max} - V_{oc}^{ex}$, $\Delta J_{sc} = J_{sc}^{max} - J_{sc}^{ex}$ and $\Delta FF = FF^{max} - FF^{ex}$, where $V_{oc}^{ex}$, $J_{sc}^{ex}$ and $FF^{ex}$ represent the corresponding experimental values. The actual values of $\Delta\eta^{Voc}$, $\Delta\eta^{Jsc}$ and $\Delta\eta^{FF}$ in Fig. 10 are summarized in Supplemental Material Table V [14].

Figure 11 further compares (a) conversion efficiency, (b) $J_{sc}$, (c) $V_{oc}$ and (d) $FF$ obtained from the QE limit calculations with those reported in the experimental record-efficiency cells. In this figure, the shaded area for the calculated $J_{sc}$ represents the $J_{sc}$ loss caused by the 5% shadow loss. The result of Fig. 11 is consistent with the data shown in Fig. 10 and Table III.

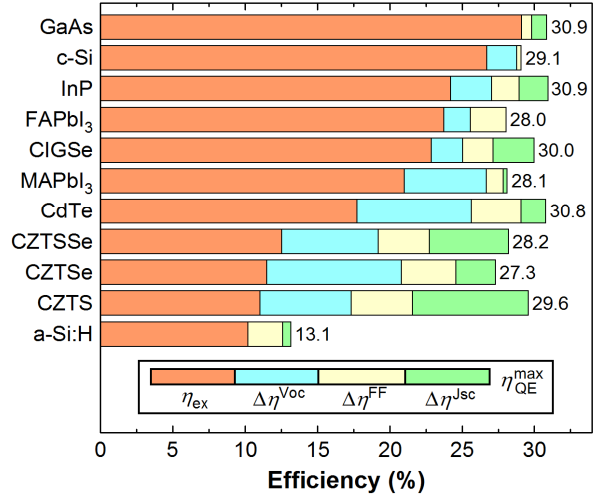

FIG. 10. Limiting factors of record-efficiency solar cells calculated from the thin-film QE limit. The experimental conversion efficiencies ($\eta_{ex}$) and the maximum efficiencies estimated from the thin-film QE limits ($\eta_{QE}^{max}$) are summarized. The difference between $\eta_{QE}^{max}$ and $\eta_{ex}$ is further categorized into the efficiency reduction owing to $V_{oc}$ deficit ($\Delta\eta^{Voc}$), $J_{sc}$ deficit ($\Delta\eta^{Jsc}$) and $FF$ deficit ($\Delta\eta^{FF}$).

Table III. Absorber thicknesses, maximum efficiencies and performance deficits ($\Delta V_{oc}$, $\Delta J_{sc}$, $\Delta FF$) of inorganic and hybrid perovskite solar cells. For the maximum efficiencies, the values obtained from the experiment ($\eta_{ex}$) and the QE limit calculation ($\eta_{QE}^{max}$) are shown.

| Solar cell | Thickness (μm) | Ref. [a] | $\eta_{ex}$ (%) | Ref. [b] | $\eta_{QE}^{max}$ (%) | $\Delta V_{oc}$ (V) | $\Delta J_{sc}$ (mAcm$^{-2}$) | $\Delta FF$ |
|---|---|---|---|---|---|---|---|---|
| GaAs | 2 | [45] | 29.1 | [5] | 30.9 | −0.012 | 1.250 | 0.025 |
| c-Si | 165 | [46] | 26.7 | [46] | 29.1 | 0.119 | −3.550 | 0.019 |
| InP | 5 | [47] | 24.2 | [5] | 30.9 | 0.102 | 2.420 | 0.060 |
| FAPbI$_3$ | 0.65 | [48] | 23.7 | [5] | 28.0 | 0.116 | −1.264 | 0.105 |
| CIGSe | 3 | [49] | 22.9 | [5] | 30.0 | 0.063 | 4.380 | 0.066 |
| MAPbI$_3$ | 0.5 | [50] | 21.1 | [50] | 28.1 | 0.267 | 0.305 | 0.045 |
| CdTe | 3.5 | [51] | 17.8 | [56] | 30.8 | 0.317 | 2.310 | 0.126 |
| CZTSSe | 2 | [52] | 12.6 | [52] | 28.2 | 0.200 | 11.450 | 0.150 |
| CZTSe | 2.2 | [53] | 11.6 | [53] | 27.3 | 0.251 | 7.560 | 0.169 |
| CZTS | 0.9 | [54] | 11.0 | [5] | 29.6 | 0.293 | 10.934 | 0.191 |
| a-Si:H | 0.22 | [55] | 10.2 | [55] | 13.1 | −0.025 | 1.010 | 0.171 |

[a] Reference for the layer thickness of each absorber. [b] Reference for the experimental record efficiency ($\eta_{ex}$) of each solar cell.

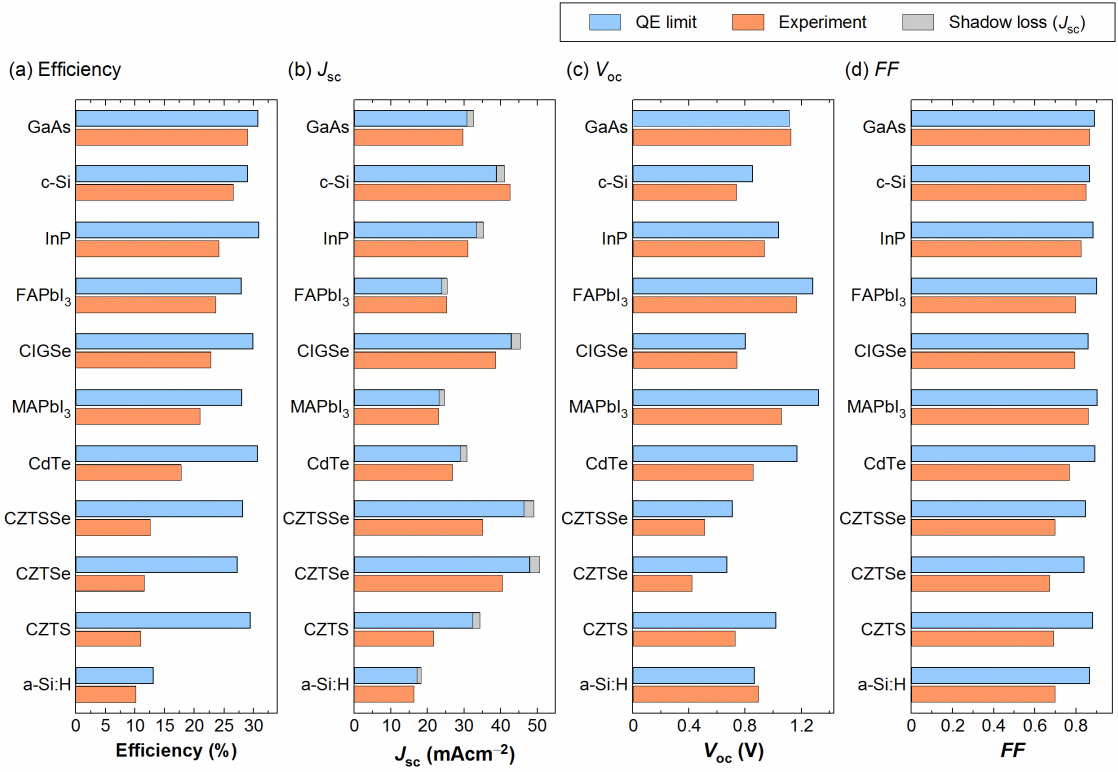

FIG. 11. (a) Conversion efficiency, (b) $J_{sc}$, (c) $V_{oc}$ and (d) $FF$ obtained from the QE limit calculations and experiment. The parameter values indicated for the QE limit calculations are consistent with Fig. 10. The experimental values of record-efficiency solar cells are adopted from Refs. [5,46,50,52,53,55,56].

For the hybrid perovskite, the highest efficiency (23.7%) is reported for $(FAPbI_3)_{1-x}(MAPbBr_3)_x$, but the calculation was performed assuming a pure $FAPbI_3$ phase as $x$ is small. As for c-Si and $FAPbI_3$-based solar cells, the experimental solar cell shows a slightly higher $J_{sc}$ than $J_{sc}^{max}$. This is attributed to the lack of the front metal-grid electrodes in the experimental cells; although the shadow loss of 5% is assumed in our calculation, in the record-efficiency c-Si and $FAPbI_3$-based solar cells, there are no front metal electrodes and $J_{sc}$ of these cells becomes higher. The presence of a large pyramid-type texture (c-Si) and a slight difference in $E_g$ ($FAPbI_3$) also contribute to increase $J_{sc}$ of the experimental cells. On the other hand, even though a CdTe-based alloy is used in the record efficiency solar cell (22.1% in Ref. [5]), the calculation result obtained for a pure CdTe solar cell (17.8% in Ref. [56]) is shown in Figs. 10 and 11 due to the uncertainty of the absorber optical properties.

Among all the experimental solar cells, the GaAs cell shows the highest conversion efficiency [5]. As shown in Figs. 10 and 11, the overall efficiency loss of the GaAs cell is quite small and the experimental efficiency is very close to the maximum limit. A slight negative $\Delta V_{oc}$ obtained for the GaAs (see Table III) is caused by the difference between the actual and assumed structures and more detailed discussion for the $V_{oc}$ loss in a GaAs cell is described in Sec. IV A.

In many other solar cells, including c-Si, hybrid perovskite, CdTe and CZT(S)Se, the main cause of the efficiency drop is attributed to $V_{oc}$ loss. In general, $V_{oc}$ can be related to $FF$ [57] and $V_{oc}$ loss tends to increase with $FF$ loss. In other words, many record-efficiency inorganic and hybrid perovskite solar cells are limited by $\Delta\eta^{Voc}$ and $\Delta\eta^{FF}$. As confirmed from Fig. 10, $\Delta\eta^{Voc}$ and $\Delta\eta^{FF}$ increase notably in polycrystalline absorbers (i.e., hybrid perovskite, CIGSe, CdTe and CZT(S)Se), suggesting the efficiency reduction by grain boundary recombination. In fact, the formation of larger polycrystalline grains has been the key for improved efficiencies in MAPbI$_3$ [58,59], CIGSe [60], CdTe [61,62] and CZT(S)Se [52,53] solar cells. However, interface recombination is also expected to contribute to $V_{oc}$ loss in these cells (see Sec. IV C for MAPbI$_3$). In CIGSe solar cells, to suppress the interface recombination, a V-shaped Ga grading structure has been incorporated [60,63] and, indeed, the CIGSe solar cell shows small $\Delta\eta^{Voc}$, compared with other polycrystalline solar cells (MAPbI$_3$, CdTe, CZT(S)Se).

One of the remarkable features of hybrid perovskite solar cells is negligible $J_{sc}$ loss. In fact, the earlier EQE analysis of the hybrid perovskite cells shows clearly that the parasitic absorption of all the component layers is very small and the efficient optical confinement is realized by metal backside reflection [24,64]. Thus, quite high efficiencies confirmed for the hybrid perovskites can be interpreted by the very low optical losses. In contrast, all the chalcogenide-based solar cells (CIGSe and CZT(S)Se) show rather large $\Delta\eta^{Jsc}$. This is mainly caused by the strong parasitic absorption in the Mo back electrode [23,24], which has been used commonly in these solar cells. In particular, the reflectivity at the semiconductor/Mo interface is quite low (~40%) [9] and strong light absorption occurs in the Mo rear metal. Thus, for the further improvement, a better optical architecture is necessary. The performance limiting factors of c-Si, hybrid perovskite (MAPbI$_3$), and a-Si:H solar cells are further discussed in Sec. IV.

Figure 12 summarizes the ratio of the experimental efficiency and maximum efficiency derived from the thin-film QE limit ($\eta_{ex}/\eta_{QE}^{max}$). In this figure, $\eta_{ex}/\eta_{QE}^{max}$ is shown by the two dimensional variables of the $J_{sc}$ ratio ($J_{sc}/J_{sc}^{max}$) and the $V_{oc}FF$ ratio ($V_{oc}FF/V_{oc}^{max}FF^{max}$). In this figure, each bar is divided into the contribution of $\Delta\eta^{Voc}$, $\Delta\eta^{FF}$ and $\Delta\eta^{Jsc}$ and, if the experimental efficiency is low, the height of the bar increases. It can be confirmed that many high-efficiency inorganic and hybrid perovskite solar cells are limited by $V_{oc}$ and $FF$. At this stage, only two solar cells (GaAs and c-Si) show suppressed overall $V_{oc}$, $FF$ and $J_{sc}$ losses.

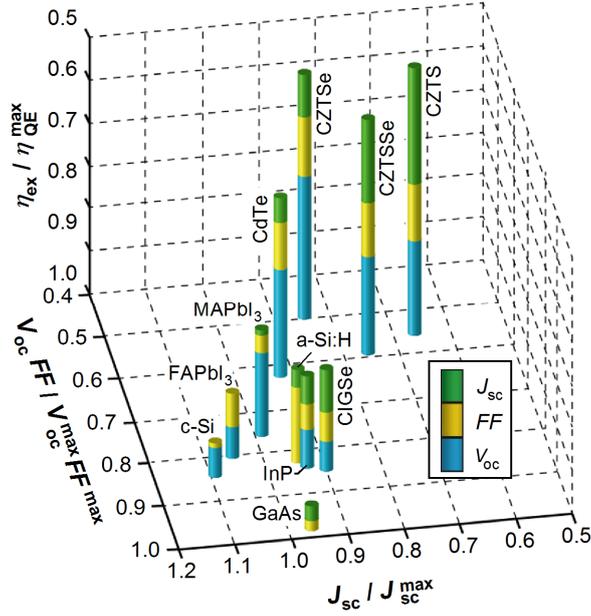

FIG. 12. Ratio of the experimental efficiency ($\eta_{ex}$) and maximum efficiency calculated from the thin-film QE limit ($\eta_{QE}^{max}$). The ratio of $\eta_{ex}/\eta_{QE}^{max}$ is shown by the two dimensional variables of the $J_{sc}$ ratio ($J_{sc}/J_{sc}^{max}$) and the $V_{oc}FF$ ratio ($V_{oc}FF/V_{oc}^{max}FF^{max}$) for each record-efficiency solar cell.

## IV. DISCUSSION

### A. GaAs solar cell

In the estimation of the maximum performance limit described above, a textured structure has been assumed, whereas flat structures have been adopted in high-efficiency GaAs solar cells. Such a change in the structural configuration modifies the longer-$\lambda$ EQE response, leading to $V_{oc}^{max}$ change. Thus, to address the true $V_{oc}$ loss in GaAs solar cells, we have further performed the EQE analysis of a practical GaAs solar cell with $\eta_{ex}$ = 27.8% ($V_{oc}$ = 1.10 V, $J_{sc}$ = 29.4 mA/cm$^2$, $FF$ = 0.857), reported in Ref. [65]. This particular GaAs cell was chosen as the exact device structure is known.

Figure 13(a) shows the EQE and $R$ spectra of the experimental GaAs cell (open circles) and the analyzed result (solid lines). The device structure is shown in the inset. In this device, the photocarriers generated in both GaAs p and n layers are collected and the sum of these contributions provides excellent fitting to the experimental EQE. Figure 13(b) indicates the $\lambda$-dependent $J_0$ contribution ($J_{0,\lambda}$) calculated from Eqs. (3) and (4). Specifically, when the incident angle of the EQE calculation is fixed to zero (i.e., $\theta$ = 0° and $R \neq 0$), $J_{0,\lambda}$ is calculated as $J_{0,\lambda}$ = (1 − $S$)$qQ(\lambda)\varphi_{BB}(\lambda)$ from Eq. (4). Thus, by integrating $J_{0,\lambda}$ of Fig. 13(b), $J_0$ is determined. Similarly, $J_{0,\lambda}$ of $\theta \neq$ 0° ($R \neq$ 0) in Fig. 13(b) shows the result calculated from Eq. (3). Often, the potential efficiency calculations are performed assuming $R$ = 0 and we also calculated the $J_{0,\lambda}$ spectrum when $R = 0$ ($\theta = 0°$). For the $J_{0,\lambda}$ calculations, $S = 0.02$ reported in Ref. [65] was adopted.

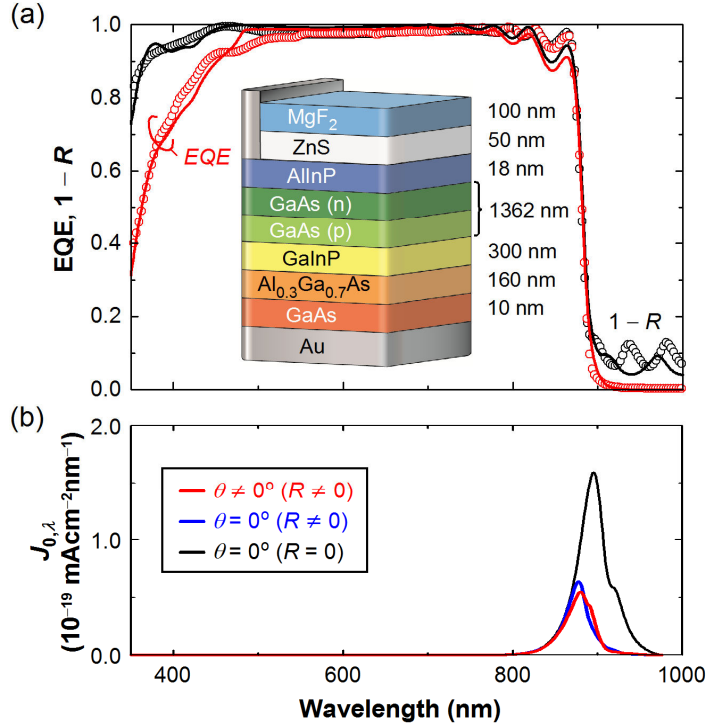

FIG. 13. (a) EQE and $R$ spectra of the experimental GaAs cell (open circles) and the simulated result (solid lines) and (b) $\lambda$-dependent $J_0$ contribution ($J_{0,\lambda}$) calculated from Eq. (3) (i.e., $\theta \neq$ 0° and $R \neq$ 0) and Eq. (4) (i.e., $\theta$ = 0° and $R \neq$ 0). The experimental spectra of (a) were taken Ref. [65]. In (b), the calculation result obtained assuming $R$ = 0 with $\theta$ = 0° is also shown.

Table IV. $J_0$ and $V_{oc}^{max}$ of the high-efficiency GaAs solar cell, obtained from the calculation of Fig. 13(b) and experiment. For the calculation, the results estimated from the EQE calculations using Eq. (3) (i.e., $\theta \neq 0°$) and Eq. (4) (i.e., $\theta = 0°$) are shown. In the simulation of $\theta = 0°$ ($R = 0$), we assumed $R = 0$ with $\theta = 0°$.

| Method | $J_0$ ($10^{-18}$ mAcm$^{-2}$) | $V_{oc}^{max}$ (V) |
|---|---|---|
| $\theta \neq 0°$ ($R \neq 0$) | 2.310 | 1.138 |
| $\theta = 0°$ ($R \neq 0$) | 2.379 | 1.137 |
| $\theta = 0°$ ($R = 0$) | 7.220 | 1.109 |
| Experiment [a] | – | 1.101 |

[a] Ref. [65].

It can be seen from Fig. 13 that $J_0$ is determined primarily in the longer-$\lambda$ region. The $J_{0,\lambda}$ values obtained assuming $\theta \neq 0°$ and $\theta = 0°$ ($R \neq 0$) are quite similar, confirming that $\theta = 0°$ is a valid assumption. In contrast, the calculation with an assumption of $R = 0$ significantly increases $J_{0,\lambda}$. Table IV summarizes $J_0$ and $V_{oc}^{max}$ obtained from Fig. 13(b) and the experimental GaAs cell. The result of Table IV shows that $\Delta V_{oc}$ of the high-efficiency GaAs cell is quite small (~40 mV), indicating quite suppressed non-radiative recombination in the cell. As confirmed from Figs. 10 and 11, the efficiency of GaAs is mainly limited by $J_{sc}$. The smaller $J_{sc}$ in the high-efficiency cells can be interpreted by unfavorable parasitic absorption mainly in the AlInP layer [$\Delta J_{sc}$ ~ 1 mA/cm$^2$ in Fig. 13(a)].

**B. c-Si solar cell**

In Fig. 10, the c-Si solar cell shows surprisingly high $\Delta \eta^{Voc}$, while $\Delta \eta^{FF}$ and $\Delta \eta^{Jsc}$ are negligible. As pointed out previously [66,67], in the indirect-transition c-Si solar cells, $V_{oc}$ deficit increases rather significantly by the Auger recombination (i.e., recombination within the bulk), while the defect-induced non-radiative recombination plays a minor role. Indeed, earlier works showed that the Auger recombination is the dominant recombination mechanism in conventional c-Si cells [66–68] and the variation of $V_{oc}$ with c-Si wafer thickness can be explained by considering only radiative (i.e., $J_0$) and Auger recombinations [68].

For c-Si solar cells, to clarify the absolute performance limit, a quite detailed calculation has been made by incorporating Coulomb-enhanced Auger recombination, $E_g$ renormalization and free carrier absorption [69]. In particular, when the optical confinement based on the Lambertian scheme is assumed, the c-Si efficiency limit of 29.4% ($V_{oc}$ = 0.761 V, $J_{sc}$ = 43.3 mA/cm$^2$, $FF$ = 0.892) has been obtained for a 110-μm-thick c-Si solar cell without considering the shadow loss. Although this performance limit is exactly the same as that shown in Table II, the limiting factors are different. Specifically, $V_{oc}^{max}$ in our calculation is ~0.86 V, while $V_{oc}$ of the above calculation is 0.761 V, which is quite close to $V_{oc}$ = 0.738 V reported for the record-efficiency c-Si solar cell [46] (i.e., $\Delta V_{oc}$ = 23 mV). Thus, among various solar cells, the c-Si solar cell is unique and its $V_{oc}$ is limited by the Auger recombination.

### C. MAPbI$_3$ solar cell

As described in Sec. III D, high efficiencies of hybrid perovskite solar cells are featured by quite low $\Delta J_{sc}$. Such suppressed $J_{sc}$ losses have been realized only in hybrid perovskite and textured c-Si solar cells and all the other thin-film solar cells show notable efficiency reduction due to $J_{sc}$ loss.

Although the MAPbI$_3$ solar cell analyzed in Figs. 10–12 is a record-efficiency cell with $V_{oc}$ = 1.06 V, $J_{sc}$ = 23.1 mA/cm$^2$, $FF$ = 0.86, $\Delta V_{oc}$ of this p-i-n-type solar cell is relatively large (0.267 V in Table III). For hybrid perovskite solar cells, however, larger $V_{oc}$ values exceeding 1.2 V have been confirmed [22,70,71]. Quite recently, for a p-i-n MAPbI$_3$ solar cell ($\eta_{ex}$ = 20.2%), a remarkable high $V_{oc}$ of 1.26 V has been obtained [22]. This $V_{oc}$ corresponds to $\Delta V_{oc}$ of only 60 mV, which can be realized by eliminating both bulk and interface recombinations.

Recently, the rather significant interface recombination in hybrid perovskite solar cells has been revealed and implied $V_{oc}$ changes notably by the choice of the electron and hole transport layers [72-74]. Specifically, as a hole transport layer (HTL), a PTAA layer provides high $V_{oc}$ with suppressed interface recombination, while the application of a PEDOT:PSS HTL reduces the implied $V_{oc}$ by 100 mV, compared with the PTAA HTL [74]. Unfortunately, in the p-i-n configuration, the PTAA layer placed on the light incident side exhibits strong parasitic absorption, reducing $J_{sc}$ of the cell [22]. Thus, even though the p-i-n solar cell with the PTAA HTL shows a record $V_{oc}$ of 1.26 V, $J_{sc}$ of this cell (~20 mA/cm$^2$) is rather limited. In contrast, the record-efficiency MAPbI$_3$ cell in a similar p-i-n configuration, which adopts the PEDOT:PSS HTL, shows higher $J_{sc}$ and lower $V_{oc}$, compared with the PTAA cell. Accordingly, at this stage, there is a trade-off between $J_{sc}$ and $V_{oc}$ in high-efficiency perovskite solar cells.

**D. a-Si:H solar cell**

Interestingly, for the disordered a-Si:H cell, our calculation shows that the experimental $V_{oc}$ is comparable to $V_{oc}^{max}$ ($\Delta V_{oc} \sim 0$). Thus, although the absolute $V_{oc}$ of a-Si:H cells is low, the experimental $V_{oc}$ of the a-Si:H cell can be explained by considering the strong tail absorption. This result is consistent with an earlier work of Tiedje who explained that $V_{oc}$ of a-Si:H solar cells is limited by the strong recombination in the a-Si:H tail region [75]. In contrast, based on detailed EQE analyses, Rau et al. reported that non-radiative recombination is the dominant mechanism of the $V_{oc}$ reduction in a-Si:H [8]. Unfortunately, the optical characteristics of a-Si:H depend strongly on process conditions [34,76]; the $E_g$ position and tail absorption change particularly by the incorporation of H. Thus, the controversy could be attributed to the difference in a-Si:H optical properties and device structures to some extent.

It should be noted that a quite thin absorber (~220 nm) is used for the record-efficiency a-Si:H cell [55], to suppress the photodegradation effect [77]. The $J_{sc}$ of the solar cell is therefore limited severely by this thin layer thickness. Consequently, the low efficiency limit observed for a-Si:H can be interpreted by low $V_{oc}$ and $J_{sc}$.

**E. Applicability of the QE limit**

As described in Sec. II, an optical model [Fig. 1(b)] and several parameters have been assumed in the evaluation of the QE limits. Here, we discuss the validity of our input parameters for the QE limit estimation. In the calculation of $J_0$ using Eq. (3), we performed the integration of the blackbody radiation with a resolution of $\Delta\theta = 0.1^o$. However, the calculated $J_0$ is rather insensitive to the integration step and shows a little change when $\Delta\theta$ is less than $10^o$. Although a fixed temperature of 300 K has been assumed in our method, the efficiency decreases rather significantly with increasing temperature. When the temperature defined in Eq. (1) is varied, the QE limit reduces with a coefficient of −0.037 %/K in the case of a GaAs cell (1 μm). Accordingly, the identical temperature needs to be employed for comparison.

In the optical model of the QE limit calculation, the presence of the antireflection layers is critical and the QE limit reduces from 30.7% to 29.7% when the dual antireflection coating is removed in the GaAs cell. In the thin-film model, a high-mobility TCO ($In_2O_3$:H) has also been assumed. If this TCO is replaced with a conventional $In_2O_3$:Sn layer with a mobility of 23 $cm^2$/(Vs) (a carrier concentration of 5 × $10^{20}$ $cm^{-3}$) [9], free carrier absorption increases, leading to the lower $J_{sc}$. This $J_{sc}$ reduction enhances in the light absorber with smaller $E_g$ as free carrier absorption increases at lower $E$ [17]. In the case of the GaAs cell, the $J_{sc}$ reduction caused by the replacement of the TCO (i.e., from $In_2O_3$:H to $In_2O_3$:Sn) is 0.7 mA/$cm^2$, whereas the $J_{sc}$ reduction increases to 2.3 mA/$cm^2$ in the CISe cell with a lower $E_g$. Thus, in low-$E_g$ absorbers, the incorporation of the high-mobility TCO becomes more important.

## V. CONCLUSION

Maximum conversion efficiencies of 13 inorganic and hybrid perovskite solar cells in a 1-μm-thick physical limit have been evaluated by extending the physical theory established by Shockley and Queisser. To determine realistic efficiency limits, a perfectly realizable thin-film solar cell structure was constructed. We performed a strict evaluation of $J_0$ by fully integrating the blackbody radiation toward the thin-film solar cell placed within a spherical cavity. In this approach, rigorous calculations for polarization- and angle-dependent quantum efficiency spectra have been implemented. In the estimation of the thin-film maximum efficiencies, the effects of absorber tail absorption and optical confinement by texturing and back-side reflection have been incorporated explicitly. In contrast to the SQ limit, which is based on unphysical assumptions of infinite absorber thickness and zero reflection, our thin-film limit calculation provides real-world efficiency limits. For the absorbers with a sharp absorption tail (GaAs, CIGSe, CdTe, InP, CZGSe), over 30% efficiencies have been confirmed in thin film form. In contrast, the absorbers with disordered phases (a-Si:H and CZT(S)Se) show deteriorated efficiencies due to the presence of the strong tail absorption, which reduces $V_{oc}$ significantly. We find that the efficiency of a record-efficiency GaAs solar cell is quite close to the maximum limit, whereas many other high efficiency cells, including hybrid perovskite solar cells, are limited primarily by $V_{oc}$ and $FF$. Our rigorous approach is quite effective in evaluating the possible maximum conversion efficiencies and limitations of record-efficiency solar cells.

# Supplementary Information

**Maximum efficiencies and performance limiting factors of inorganic and hybrid perovskite solar cells**

Yoshitsune Kato, Shohei Fujimoto, Masayuki Kozawa and Hiroyuki Fujiwara

*Department of Electrical, Electronic and Computer Engineering, Gifu University, 1-1 Yanagido, Gifu 501-1193, Japan.*

**Contents:**

1. Optical constants of component layers.
2. Supporting tables and figures.

## 1. Optical constants of solar-cell component layers used in thin-film limit calculation

The optical constants (refractive index $n$ and extinction coefficient $k$) of the solar cell component layers ($MgF_2$, $Al_2O_3$, $In_2O_3$:H, Ag) adopted in the thin film model shown in Fig. 1(b) are summarized in Supplementary Figure 1. All these optical functions were calculated from the optical constant models; the $n$ spectra of $MgF_2$ and $Al_2O_3$ were derived from the Sellmeier model, whereas the ($n$, $k$) spectra of $In_2O_3$:H and Ag were estimated by combining the Tauc-Lorentz model with the Drude model.

The expression of the Sellmeier model is given by

$$n^2(\lambda) = \frac{B_1 \lambda^2}{(\lambda^2 - C_1)} + \frac{B_2 \lambda^2}{(\lambda^2 - C_2)} + \frac{B_3 \lambda^2}{(\lambda^2 - C_3)} + 1, \quad \text{(S1)}$$

where $B$ and $C$ represent the model parameters.

For the ($n$, $k$) calculation of $In_2O_3$:H and Ag, the $\varepsilon_2$ spectrum ($\varepsilon_2 = 2nk$) was calculated first according to

$$\varepsilon_2(E) = \sum_{j=1}^{m} \frac{A_j C_j E_{0,j}(E - E_{g,j})^2}{[(E^2 - E_{0,j}^2)^2 + C_j^2 E^2]E} + \frac{A_D \Gamma}{E^3 + \Gamma^2 E} . \tag{S2}$$

In this equation, the first term shows the Tauc-Lorentz model [1,2] and the second term indicates the Drude model. The Tauc-Lorentz peak is expressed by four parameters: i.e., the amplitude parameter $A_j$, broadening parameter $C_j$, peak transition energy $E_{0,j}$, and optical gap $E_{g,j}$ of the $j$th Tauc-Lorentz peak. The Drude model parameters of $A_D$ and $\Gamma$ can further be related to the optical carrier concentration ($N_{opt}$) and optical mobility ($\mu_{opt}$) [3].

Using the parameters of Eq. (S2), $\varepsilon_1(E)$ can then be calculated from the Kramers-Kronig integration of each Tauc-Lorentz peak and the $\varepsilon_1$ contribution of the Drude term:

$$\varepsilon_1(E) = \sum_{j=1}^{m} \left( \varepsilon_{1,j}(\infty) + \frac{2}{\pi} P \int_{E_g}^{\infty} \frac{E' \varepsilon_{2,j}(E')}{E'^2 - E^2} dE' \right) - \frac{A_D}{E^2 + \Gamma^2} , \tag{S3}$$

where $\varepsilon_{1,j}(\infty)$ represents a constant contribution to $\varepsilon_1(E)$ at high energies.

The Sellmeier parameters of $MgF_2$ and $Al_2O_3$ and the Tauc-Lorentz/Drude parameters of the $In_2O_3$:H and Ag were summarized in Supplementary Tables II ~ IV. These values were adopted from Ref. [4]. For the $In_2O_3$:H, the Drude parameters ($A_D$, $\Gamma$) were obtained based on the reference experimental values of $N_{opt} = 2 \times 10^{20}$ cm$^{-3}$ and $\mu_{opt} = 100$ cm$^2$/(Vs), reported in Ref. [5].

Supplementary Table I. MgF$_2$ and Al$_2$O$_3$ layer thicknesses used for each cell. The MgF$_2$ and Al$_2$O$_3$ layer thicknesses of each cell were optimized to maximize $J_{sc}$ of the solar cell.

| Solar cell | MgF$_2$ (nm) | Al$_2$O$_3$ (nm) |
| --- | --- | --- |
| CZTSe | 398 | 96 |
| ClSe | 305 | 381 |
| c-Si | 11 | 359 |
| CZTSSe | 171 | 72 |
| CIGSe | 0 | 40 |
| CZTS | 130 | 136 |
| InP | 6 | 32 |
| CZGSe | 195 | 72 |
| GaAs | 0 | 244 |
| CdTe | 134 | 98 |
| FAPbI$_3$ | 96 | 43 |
| MAPbI$_3$ | 153 | 100 |
| a-Si:H | 149 | 9 |
| CGSe | 98 | 257 |

Supplementary Table II. Sellmeier parameters of MgF$_2$ and Al$_2$O$_3$ for Eq. (S1).

| Material | $B_1$ | $B_2$ | $B_3$ | $C_1$ (μm$^2$) | $C_2$ (μm$^2$) | $C_3$ (μm$^2$) |
| --- | --- | --- | --- | --- | --- | --- |
| MgF$_2$ | 0.896 | 0.699 | – | 6.73 × 10$^{-3}$ | 127.462 | – |
| Al$_2$O$_3$ | 1.125 | 0.555 | 6.747 | 0.00242 | 0.0214 | 550.747 |

Supplementary Table III. Tauc-Lorentz (TL) and Drude parameters of $In_2O_3$:H for Eqs. (S2) and (S3).

|  | $A/A_D$ (eV) | $C/\Gamma$ (eV) | $E_0$ (eV) | $E_g$ (eV) | $\varepsilon_1(\infty)$ |
|---|---|---|---|---|---|
| TL peak1 | 45.989 | 0.684 | 4.061 | 3.714 | 2.236 |
| TL peak2 | 25.507 | 1.971 | 4.162 | 3.054 | 0 |
| TL peak3 | 27.441 | 4.016 | 5.892 | 3.285 | 0 |
| TL peak4 | 10.891 | 0.436 | 7.095 | 3.069 | 0 |
| Drude | 0.994 | 0.042 | – | – | – |

Supplementary Table IV. Tauc-Lorentz (TL) and Drude parameters of Ag for Eqs. (S2) and (S3).

|  | $A/A_D$ (eV) | $C/\Gamma$ (eV) | $E_0$ (eV) | $E_g$ (eV) | $\varepsilon_1(\infty)$ |
|---|---|---|---|---|---|
| TL peak1 | 3.671 | 2.344 | 1.179 | 0.427 | 1.928 |
| TL peak2 | 1.305 | 1.904 | 2.381 | 0.863 | 0 |
| TL peak3 | 0.162 | 0.700 | 3.339 | 0.217 | 0 |
| TL peak4 | 233.000 | 0.572 | 3.949 | 3.748 | 0 |
| TL peak5 | 85.379 | 1.825 | 4.130 | 3.766 | 0 |
| TL peak6 | 23.454 | 1.256 | 5.322 | 4.840 | 0 |
| Drude | 74.909 | 0.039 | – | – | – |

Supplementary Table V. Optimized $MgF_2$ and $Al_2O_3$ layer thicknesses in devices and performance limitations due to $V_{oc}$ ($\Delta\eta^{Voc}$), $J_{sc}$ ($\Delta\eta^{Jsc}$) and FF ($\Delta\eta^{FF}$).

| Solar cell | $MgF_2$ (nm) | $Al_2O_3$ (nm) | $\Delta\eta^{Voc}$ (%) | $\Delta\eta^{Jsc}$ (%) | $\Delta\eta^{FF}$ (%) |
|---|---|---|---|---|---|
| GaAs | 2 | 34 | −0.30 | 1.22 | 0.83 |
| c-Si | 270 | 125 | 3.64 | −1.90 | 0.61 |
| InP | 6 | 34 | 2.85 | 2.03 | 1.89 |
| $FAPbI_3$ | 235 | 178 | 2.25 | −0.96 | 3.04 |
| CIGSe | 644 | 452 | 2.11 | 2.85 | 2.09 |
| $MAPbI_3$ | 282 | 112 | 5.60 | 0.28 | 1.17 |
| CdTe | 20 | 72 | 7.83 | 1.73 | 3.42 |
| CZTSSe | 584 | 515 | 6.60 | 5.48 | 3.52 |
| CZTSe | 40 | 375 | 9.23 | 2.74 | 3.79 |
| CZTS | 316 | 737 | 6.32 | 8.02 | 4.22 |
| a-Si:H | 12 | 27 | −0.26 | 0.63 | 2.55 |

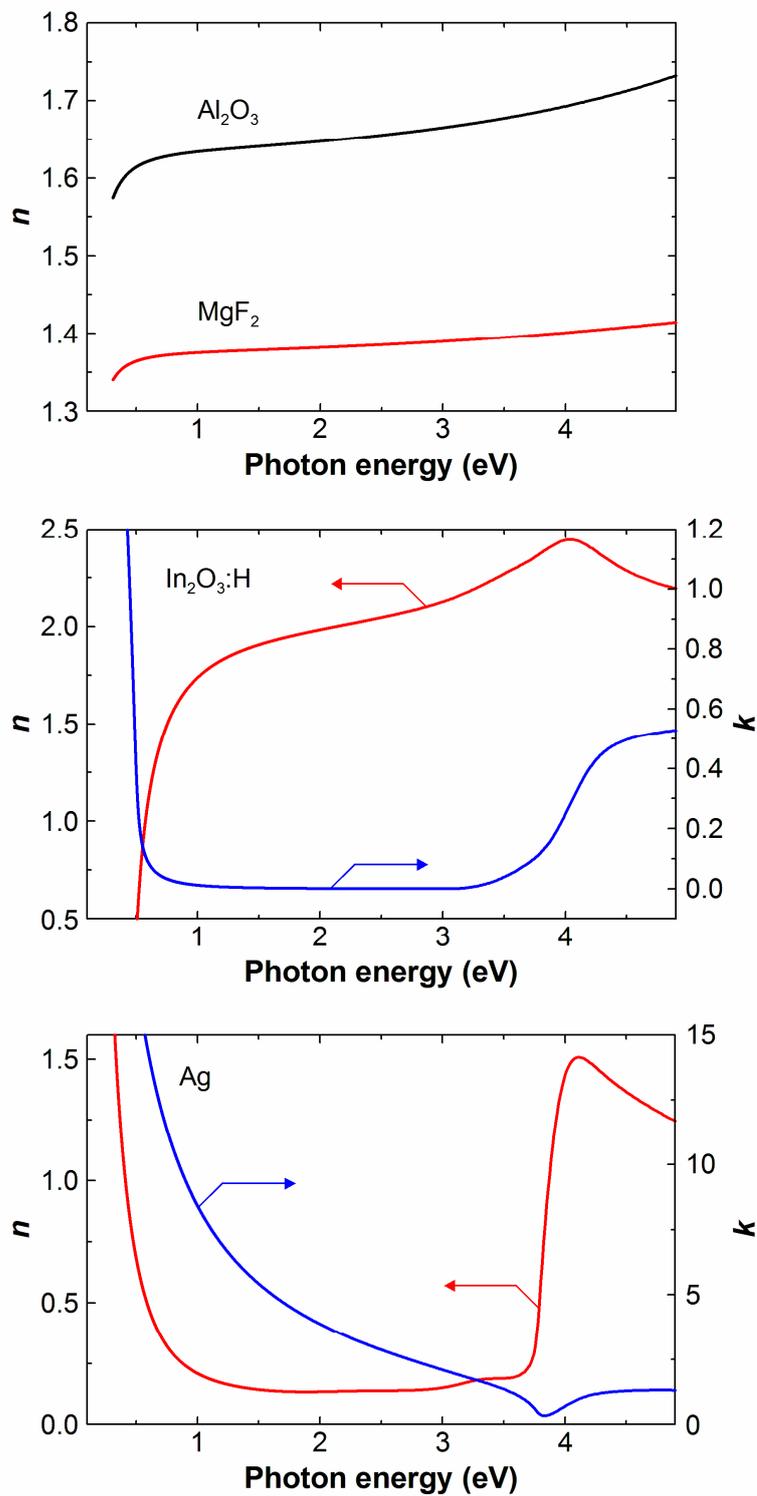

Supplementary Figure 1. Optical constants of MgF$_2$, Al$_2$O$_3$, In$_2$O$_3$:H and Ag adopted in the calculation of the thin-film QE limit. The parameter values of these spectra are summarized in Supplementary Tables II~IV.

**References (Supplementary Information)**